\documentclass[fleqn,twoside]{article}
\usepackage{espcrc2}

\usepackage{graphicx}
\usepackage[figuresright]{rotating}

\newcommand{\AmS}{{\protect\the\textfont2
  A\kern-.1667em\lower.5ex\hbox{M}\kern-.125emS}}
\newcommand{\numutau}{\nu_{\mu}\rightarrow\nu_{\tau}}
\newcommand{\numunue}{\nu_{\mu}\rightarrow\nu_{e}}
\newcommand{\nutau}{\nu_{\tau}}
\title{The CERN-Gran Sasso Neutrino Program}
\author{D. Duchesneau\address{LAPP,
        IN2P3-CNRS, \\
        Chemin de Bellevue, BP110, F-74941, Annecy-le-Vieux, France}%
        \thanks{Representing the OPERA and ICARUS Collaborations.}}

\begin{document}

\begin{abstract}
This paper reviews the current experimental
program envisaged with the future CERN neutrino
beam called CNGS. Two detectors, OPERA and ICARUS, are under preparation
and should investigate the neutrino properties coming from the CNGS to shed
 light on neutrino oscillation physics.

\vspace{1pc}
\end{abstract}

\maketitle

\section{Introduction}

Among the various recent experimental results concerning neutrino physics,
two of them give very strong hints for the existence of an oscillation 
mechanism and they are driving the main motivations for the future projects.
The first one is the clear $\nu_{\mu}$ disappearance observed by 
Super-Kamiokande, SoudanII 
and Macro in the atmospheric data which can be very well
described by the presence of $\nu_{\mu} \rightarrow \nu_{\tau}$ oscillation.
The best fit of this hypothesis to the latest Super-Kamiokande data gives 
the oscillation parameters $\Delta m^{2}=2.5$x$10^{-3}$$ eV^{2}$ 
with $sin^{2}2\theta$=1.0~\cite{SK}.
The range of allowed values at 90\% CL correspond to  
1.6x$10^{-3} < \Delta m^{2} <$ 3.9x$10^{-3}$$ eV^{2}$ and $sin^{2}2\theta>$0.92.

The second result concerns the well established  solar neutrino deficit 
for which a clear flavour change process has been demonstrated with the 
recently published SNO
data~\cite{SNO}. The solar observations are
compatible with an oscillation process favouring
again a large mixing angle but at a $\Delta m^{2}$ one order of magnitude less
than for the atmospheric data. 

At this stage it is very important to 
test in a conclusive manner an oscillation mechanism or not
as the origin of those results.
The primary  goal of the first generation of  Long Baseline projects
is to  confirm and verify the nature of the oscillations 
observed in the atmospheric data as well as to provide
 more precise measurements of the corresponding oscillation parameters.
Three projects using `home-made' $\nu_{\mu}$ are under progress. The 
K2K~\cite{K2K} and
Numi/Minos~\cite{minos} projects are looking primarily at  $\nu_{\mu}$ disappearance using
low energy beams,
while the main goal of the CNGS project described in this paper
 is to search for
$\nu_{\tau}$ appearance in a high energy $\nu_{\mu}$ beam at 730 km from 
the neutrino source. 

\section{The CNGS beam line}
\begin{figure}[htb]
\vspace{-0.3cm}
\begin{center}
    \includegraphics[width=7cm]{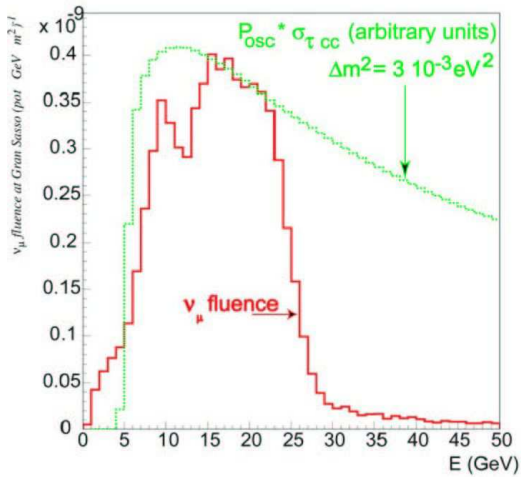}
\end{center}
\vspace{-1.5cm}
\caption{CNGS $\nu_{\mu}$ flux as a function of the energy
 at Gran Sasso. The curve shows the energy dependence of the  production rate
of  $\tau$'s produced via charged current interaction for
 $\Delta m^{2}=$3x$10^{-3}$$ eV^{2}$ at full mixing.}
\label{fig:ebeam}
\vspace{-0.7cm}
\end{figure}

The CNGS~\cite{cngs} is a $\nu_{\mu}$ beam produced with 400 GeV protons extracted from
the SPS complex at CERN. 
During one year in a mode 
where the use of the SPS is shared  with  LHC operation, 
4.5x$10^{19}$ protons on target (pot) can be delivered, assuming 200 days of operation.
 The protons hit a target made of
graphite rods and  the produced secondary particles pass through a
magnetic focusing system  designed to select high energy (20-50 GeV) 
$\pi^{+}$ and $K^{+}$. This makes  the neutrino beam  a high
energy beam optimised
for $\nu_{\tau}$ appearance with a  mean neutrino energy of about 17 GeV.
Fig.~\ref{fig:ebeam} shows the expected neutrino  energy distribution 
at Gran Sasso  located at 732 km from CERN.
In the Gran Sasso underground laboratory, the 
$\nu_{\mu}$ flux should correspond to  3.5 x $10^{11}$ $\nu /m^{2}$/year with 
a  contamination of 2.1\% $\overline{\nu}_{\mu}$,
 0.8\% $\nu_{e}$ and less than 0.05\% of $\overline{\nu}_{e}$.
The number of charged current interactions expected from $\nu_\mu$
is about 2600 /kton/year.
If the $\numutau$ oscillation hypothesis is confirmed 
the number of $\tau$'s produced via
charged current interaction at the Gran Sasso is about 15 /kton/year 
for  $\Delta m^{2}=$2.5x$10^{-3}$$ eV^{2}$ at full mixing. 

Civil engineering work is progressing well and it should be finished by 
 spring 2003. The accelerator team 
plans to deliver the first
neutrino beam by May 2006.
The possibility of an increase of the neutrino
beam intensity by a factor 1.5 is under study~\cite{cngs1.5}.

\section{Experimental signature}
The $\nutau$ appearance search is based on the observation
of events produced by  charged current interaction (CC) with the
$\tau$ decaying in all possible decay modes.
Since the expected event rate is small, it is crucial to separate 
efficiently the $\nutau$ CC
events from all the other flavour neutrino events and to
keep the background at a very low level.
For this purpose the detectors will have to identify the events by
exploiting the $\tau$ specific properties characterised by a non negligible lifetime
and the presence of missing transverse momentum due to the final
state $\nutau$ produced in the $\tau$ decays.

The two proposed detectors, OPERA~\cite{opera} and ICARUS~\cite{icarus},
 are using two 
different approaches
for identifying the events. The choice made by OPERA is to 
observe the $\tau$ decay topology in nuclear emulsions, while ICARUS 
will separate the $\nutau$ CC events from the background 
through kinematical criteria using a large volume  TPC filled 
with liquid argon.
 The ICARUS and OPERA  detectors will be installed in   hall B and hall C, respectively, 
of the Gran Sasso underground
laboratory. The 2400 meters of rock above the experimental halls provide
a very efficient cosmic ray shielding.\\
The detectors  are described in more details in the following
two sections. 

\section{The OPERA experiment}

\begin{figure}[htb]
\vspace{-0.8cm}
\begin{center}
    \includegraphics[width=7cm]{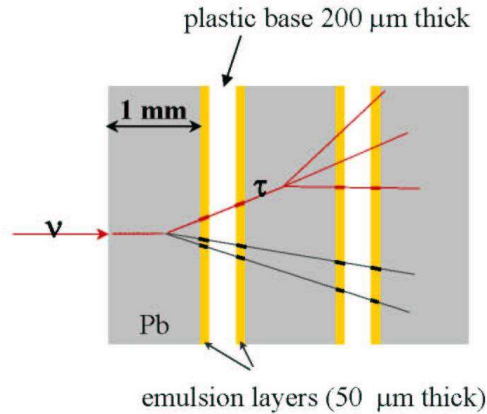}
\end{center}
\vspace{-1.5cm}
\caption{Schematic structure of an ECC cell. The $\tau$ decay kink is
reconstructed by using four track segments in the emulsion films.}
\label{fig:brick}
\vspace{-0.7cm}
\end{figure}

The principle of the OPERA experiment is to observe the $\tau$ trajectories 
and the decay products in thin layers of emulsion.
To provide the large target mass (1.8 ktons) the  emulsion films
are interleaved with  1 mm  thick lead plates. Fig.~\ref{fig:brick}
sketches the basic structure of the detector, called Emulsion Cloud Chamber (ECC).
An   emulsion film in OPERA consists of two emulsion layers (50 $\mu$m thick)
put on either side of a plastic base (200 $\mu$m thick). 

\subsection{The detector structure}
The basic detector unit, called ECC brick, is obtained by stacking 56
lead plates and emulsion films, plus an extra film before and another one,
called Changeable Sheet (CS),
behind  after 2mm of plastic. 
The CS can be detached from the rest of the brick for analysis. It will be used
  to first locate the tracks produced in neutrino interactions which
have to  be followed in the rest of the brick. 
The dimensions of a brick are  12.5 x 10.2 x 7.5 $cm^{3}$. In terms
of radiation length, a brick corresponds
to a thickness of 10 $X_{o}$. 

In order to reach  1.8 kton target mass, 206336  bricks will be 
installed into walls containing 64 rows of 52 bricks and separated
from each other by vertical planes of electronic target trackers.

\begin{figure}[htb]
\vspace{-0.8cm}
\begin{center}
    \includegraphics[width=8cm]{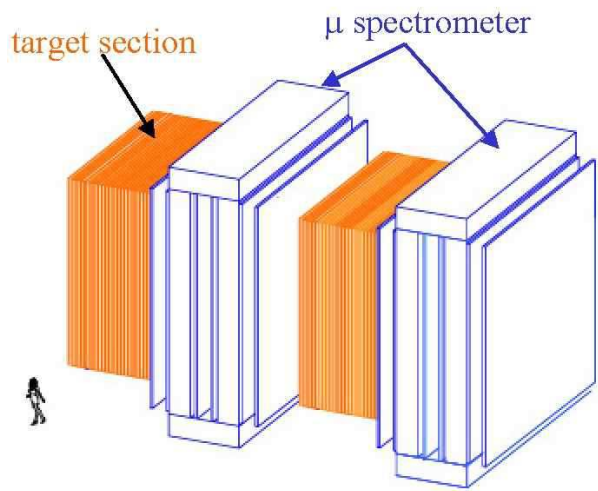}
\end{center}
\vspace{-1.5cm}
\caption{View of the OPERA detector composed of 2 super modules with 2 spectrometers.}
\label{fig:opera2SM}
\vspace{-0.7cm}
\end{figure}

 Fig.~\ref{fig:opera2SM} shows the general layout of the OPERA detector.

 It consists
of 2 identical parts called super module (SM). Each SM has a target section and
a muon spectrometer. The spectrometer measures the charge and momentum
of muons going through by  means of  a 
dipolar magnet providing 1.6 Tesla transverse to the 
neutrino beam axis and equipped with drift tubes and RPC chambers.

The target section is composed of 31 walls of bricks. The bricks will be installed
in a support structure and manipulated from the sides of the walls
using an automated manipulator.

An electronic target tracker module composed of
2 planes of  6.6 m long scintillator strips in the two transverse
directions (X and Y) is installed behind each brick wall. The main goal
of the electronic detector is to provide a trigger for the
neutrino interactions and a localisation of the brick where the event
occured. The strips, 2.6 cm wide and 1 cm thick, have WLS fibers for readout
by 64 channel multi-anode  photomultiplier tubes.
 The brick transverse pointing accuracy is about 1.5 cm for
 CC events and 3.0 cm
for NC events. The efficiency to find the right brick is about 70-80\%.

The candidate brick is then removed for subsequent analysis. 
The analysis flow is the following:
The brick is removed using the brick manipulator system and the changeable
sheet is detached and  developped. The film is then scanned to 
search for the track originating from the neutrino interaction. If none
are found then the brick is left untouched and another
one is removed. When a neutrino 
 event is observed, the brick is exposed to cosmics
to collect alignment tracks before going to the development. After
development
the emulsions are sent to the automatic scanning microscope
in order to start the analysis which consists of finding the neutrino vertex
and the decay kink in the vertex region.  

\subsection{Detector performance}
The angular resolution for reconstructing
a  track in a film  from the two emulsion layers  is about 2 mrad. This
is entirely limited by the scanning accuracy of the microscope stage and the
digitisation.
However this resolution allows a momentum measurement using the 
particle multiple scattering occuring within the lead plates. The difference
of the track angles before and after lead sheets can be measured
accurately all along the track. 
Test beam results have shown that a
 resolution better than 20\% for momentum below 4 GeV can be achieved using
only half a brick (5 $X_{o}$) with this angular method.

\begin{figure}[htb]
\vspace{0cm}
\begin{center}
    \includegraphics[width=8cm]{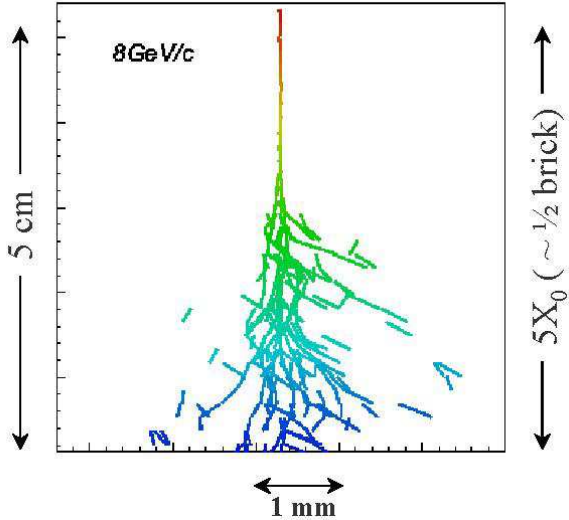}
\end{center}
\vspace{-1.5cm}
\caption{Electromagnetic shower observed in a brick exposed to an electron beam.}
\label{fig:shower}
\vspace{-0.7cm}
\end{figure}

The  brick thickness ($10 X_{o}$) and the high precision in reconstructing
track segments makes the brick a good electromagnetic calorimeter.
Fig.~\ref{fig:shower} shows an electron shower reconstructed in a brick
exposed to an electron beam at CERN. The shower structure is clearly 
visible with all the segmets representing electron tracks from the
shower seen in the emulsions. 
The efficiency to identify such showers is about 90\%.

The shower energy can be measured by counting the number of track segments 
reconstructed in a cone of 50 mrad around the incoming electron
direction. This technique gives an energy  resolution of 40\%/$\sqrt{E}$.

These results show the high  capability of OPERA in $\tau\rightarrow e$ decay studies and
in the search for $\numunue$ appearance.

\subsection{Physics performance: $\numutau$ search}
The $\tau$ decay channels investigated by OPERA are the e, $\mu$ and hadron. They
are classified  in 2 categories:  long  and short decays. The
latter corresponds to the cases where the $\tau$ decays in the
same lead plate as the neutrino interaction occured. Those events are 
selected on the basis of the impact parameter of the $\tau$ daughter
track with respect to the interaction vertex (IP $>$ 5-20 $\mu$m).
  This is applied only  for the electron
and muon channels. In the long decay category the $\tau$ does not
decay in the same lead plate and its track can be reconstructed in 
one film. The $\tau$ candidate  events are selected on the basis of the existence of
a kink angle between the $\tau$ and the daughter tracks ($\theta_{kink}>$ 20 mrad).

Table~\ref{tab:opera} summarises the OPERA  performance after 5 years of running
with the CNGS (2.25x$10^{20}$ pot). The  number of expected signal events from 
$\numutau$ oscillations is given as a function of the studied channel for three different 
values of  $\Delta m^{2}$ at full mixing.
 The total efficiency including the branching ratios amounts to 9.1\% and the
 total background is estimated
to be less than 0.65 event. The main background sources are 
 charm decays, large angle
muon scattering and hadron reinteractions.
 A 4$\sigma$ significance is achieved
after 5 years for $\Delta m^{2}>2.0$ x $10^{-3}$$eV^{2}$.
\begin{table*}[hbt]
\center
\caption{Summary of the expected numbers of $\tau$ events in 5 years for
 different $\Delta m^{2}$ with the expected background and detection efficiencies per 
decay channel for OPERA.}
\label{tab:opera}
% \begin{tabular}{|c|c|c|c|c|c|}
\begin{tabular}{@{}ccccccc}
\hline
channel  & \multicolumn{3}{c}{signal  for $\Delta m^{2}$  ($\rm{eV}^{2}$)}  &  $\epsilon$xBr & Background\\
       & 1.6x$10^{-3}$   & 2.5x$10^{-3}$ & 4.0x$10^{-3}$   &  &  \\
 \hline
$\tau\rightarrow$e      &  1.6 & 3.9 & 9.9 & 3.4\% & 0.16 \\
$\tau\rightarrow$$\mu$  &  1.3 & 3.2 & 8.2 & 2.8\% & 0.29 \\
$\tau\rightarrow$h      &  1.4 & 3.2 & 8.2 & 2.9\% & 0.20 \\
\hline
Total  &  4.3 & 10.3 & 26.3 & 9.1\% & 0.65 \\
\hline
\end{tabular}
\end{table*}

\begin{figure}[htb]
\vspace{-1cm}
\begin{center}
    \includegraphics[width=8cm]{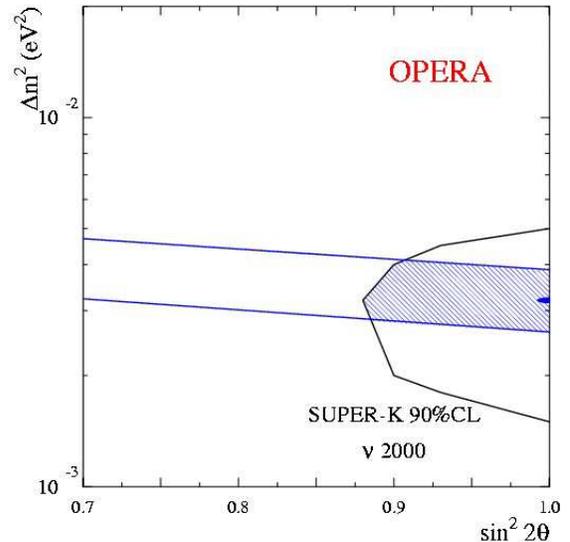}
\end{center}
\vspace{-1.5cm}
\caption{The band represents the  90\% CL allowed region for the oscillation parameters
determined by OPERA if the observed number of $\tau$ events corresponds to the expectation
for full mixing and $\Delta m^{2}$=3.2x$10^{-3} eV^{2}$.}
\label{fig:operacontour}
\vspace{-0.7cm}
\end{figure}

Fig.~\ref{fig:operacontour} shows the constraint on the value of
$\Delta m^{2}$ which can be achieved after 5 years if the
observed rate of $\nu_{\tau}$ events corresponds
to the expectation for full mixing and  $\Delta m^{2}$=3.2x$10^{-3} eV^{2}$.
 Using the atmospheric
result to constrain the mixing angle the precision 
on   $\Delta m^{2}$ at 2.5x$10^{-3}$ $eV^{2}$
is about 16\%.\\
Improvement in the background reduction is under progress.
A preliminary analysis has shown that   low energy muons
can be identified in the bricks,  using the dE/dX measurement in the
emulsions (grain density) as a function
of the particle range.
Applying this idea   reduces the estimated background from 0.65
to 0.42 without changing the signal efficiency.

\section{The ICARUS experiment}
The principle of ICARUS is to reconstruct the event kinematics with enough precision
to allow the selection of   interesting candidates and the
rejection of background events
with kinematical criteria based on energy measurements and on very good particle ID.
The technology relies on the possibility to do 3 dimensional imaging of events in a large
time projection chamber filled with liquid argon, with space resolution similar to that 
 in bubble chambers, but with electronic readout and continuous
sensitivity.
An important characteristic of this detector concept  is the use of very pure liquid argon,
with less than  1 ppb of contaminant, allowing the electrons to 
drift along distances larger than 1.5 m.
The scintillation light produced by the ionising particles in liquid argon
 is used to give a
precise reference time of the ionisation track.
The readout is achieved with 3 parallel 
wire planes (2 induction and 1 collection planes) 
with 3 different orientations, at 0,  $60^{o}$ and $-60^{o}$.
The wire pitch is 3 mm. The spatial resolution is 250$\mu$m along the
drift direction (z) and 1 mm for the x and y directions. 

In addition, the energy deposited by an ionising particle along its path (dE/dx) 
is measured accurately from the 
charge collected on each  wire of the collection plane with a time sampling of 400 nsec.
Since the particle momentum can be measured from the range of stopping
particles or from multiple scattering
measurement, it provides a very clean method for particle identification
 of soft particles.
 
\subsection{Detector structure}
\begin{figure}[htb]
\vspace{-0.5cm}
\begin{center}
    \includegraphics[width=7cm]{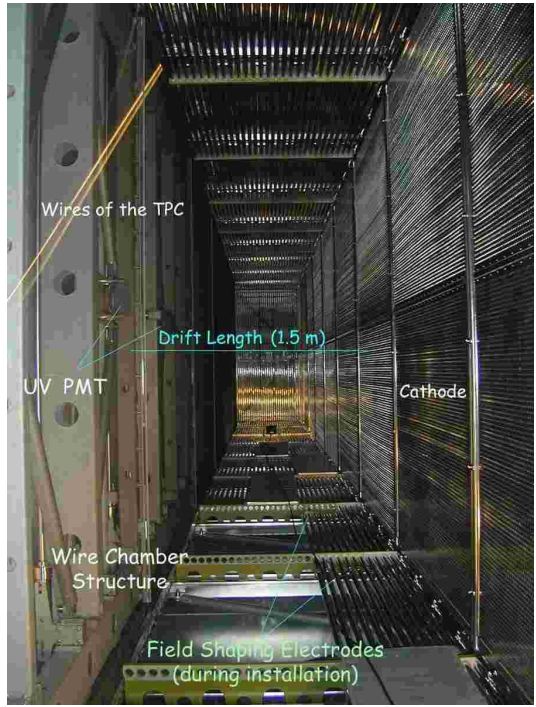}
\end{center}
\vspace{-1.cm}
\caption{Internal view of the T600 first half module.}
\label{fig:t300}
\vspace{-0.7cm}
\end{figure}

The ICARUS detector has a modular concept which should allow to build a multi kton device by 
replicating the basic component. 
The smallest detector unit contains 300 tons of liquid argon and corresponds to half a T600
module. Figure~\ref{fig:t300} shows the internal view of the first T600 half module. The
cryostat has dimensions of 4x4x20 $m^{3}$. A high voltage system produces a uniform electric
field, perpendicular to the wire planes allowing the drift of the ionisation electrons 
towards the wires over a maximum path of 1.5 m.

\begin{figure*}[hbt]
\vspace{-0.5cm}
\begin{center}
    \includegraphics[width=14cm]{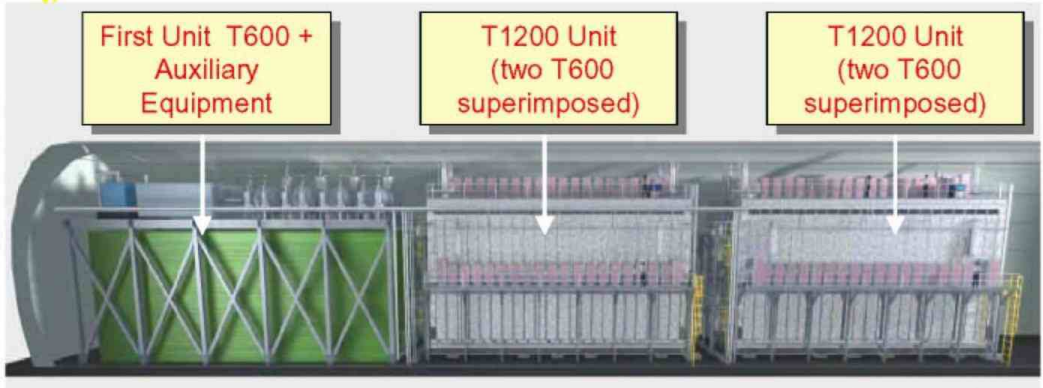}
\end{center}
\vspace{-1.0cm}
\caption{General view of the ICARUS T3000 detector composed of 5 T600 modules.}
\label{fig:t3000}
\vspace{-0.cm}
\end{figure*}          

This first module has been fully instrumented and 
succesfully tested during a technical run lasting 5 months during 2001 in Pavia (Italy).
The results were found to be in agreement with the expectations and have validated the LAr
TPC technology at these large scales.

The aim of the ICARUS collaboration is to build a 3000 ton detector (T3000) which should be
 installed in the Gran Sasso Hall B.
To fully exploit the know-how acquired with the first half T600 construction and
running,
 the T3000 design is based on cloning the T600 module to
reach a sensitive mass of 2.35 ktons of liquid argon. 
Fig.~\ref{fig:t3000} shows the general layout of the T3000 detector in the Gran Sasso
laboratory. 

The first part to be installed is the existing T600 module composed of the tested  
half module prototype
and a second half module under construction. The complete module should be installed  early 
2003 with the aim of
 starting the collection of atmospheric and solar neutrino events in 2003.
 Then the mass
should be increased gradually with time with the installation of two T1200 units (1 T1200 module is
made of 2 T600 modules superimposed in the same
 insulation enveloppe) before summer 2006.

The ICARUS physics program is quite broad. It should cover, in addition to the CNGS program,
the study
of  solar, atmospheric and supernovae  neutrino as well as  proton decay search.

\subsection{Detector performance}
The  large homogeneous tracking medium  allows
to fully identify and sample electromagnetic and hadronic
showers. The total shower energy is obtained from integrating the charge collected on the wires.
As a result ICARUS has  excellent calorimeter capabilities with very good energy resolution, 
featuring excellent electron identification and e/$\pi^{o}$ separation in addition
  to its high capability of dE/dx measurements.

Since there is no magnetic field,  momentum for energetic muon is measured using  multiple
coulomb scattering in  liquid argon. The resulting  momentum resolution is 
about 20\% for 10 GeV muons.

\subsection{Physics performance: $\numutau$ search}
 
\begin{figure}[htb]
\vspace{-0.5cm}
\begin{center}
    \includegraphics[width=7.5cm]{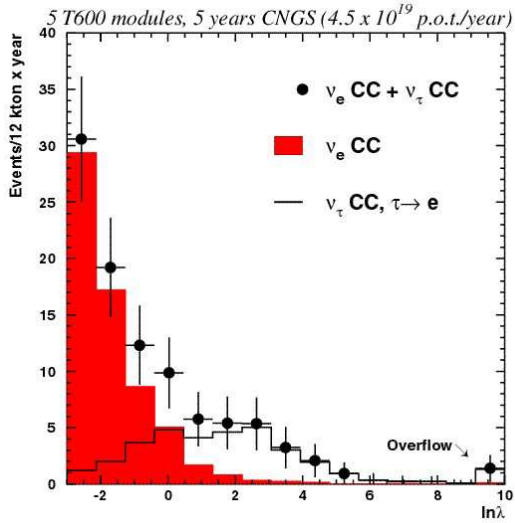}
\end{center}
\vspace{-1.5cm}
\caption{Distribution of the multi-dimensional likelihood for $\nu_{e}$ CC and $\tau\rightarrow e$
events.}
\label{fig:like}
\vspace{-0.7cm}
\end{figure}

ICARUS has considered two $\tau$ decay channels for the $\numutau$ search.
 Their ``golden'' 
channel corresponds to $\tau\rightarrow e\nu_{e}\nu_{\tau}$. The main
background should come from the $\nu_{e}$ present in the beam which interact
via charged current.
It is suppressed kinematically exploiting the fact that  signal events have missing
momentum due to the final state neutrinos which is not the case for the background events.
An analysis based on a 3 dimensional likelihood built for the signal and background events
with 3 distributions derived from the
visible energy and  transverse missing momentum distributions 
has been used. Fig.~\ref{fig:like} shows the ratio of the
corresponding signal and background likelihoods for background (shaded area)
 and $\tau$ events normalised
to 5 years of CNGS in shared mode. The two  components are clearly distinct and a cut on this
variable allows the selection of a rather clean signal sample.

The other channel which has been exploited is the $\tau\rightarrow\rho$ where
 $\rho^{-}\rightarrow\pi^{-}\pi^{o}$. The main background comes from $\nu_{\mu}$ NC events
and in this case, both signal and background events have missing energy. The selection is
then based on some isolation criteria instead of energy balance since the candidate hadron is more
isolated with respect to the total momentum for the signal.
Table~\ref{tab:icarus} summarises the overall ICARUS performance after 5 years of running
with the CNGS (2.25x$10^{20}$ pot) for $\numutau$ search. 
The  number of expected signal events from 
$\numutau$ oscillations is given as a function of the studied channel for three different 
values of  $\Delta m^{2}$ at full mixing.
 The total efficiency including the branching ratios amounts to 5.9\% and the
 total background is estimated
to be less than 0.7.
These results show that ICARUS can reach a sensitivity comparable to
 OPERA for this study.
\begin{table*}[hbt]
\center
\caption{Summary of the expected numbers of $\tau$ events in 5 years for
 different $\Delta m^{2}$ with the expected background and detection efficiencies per 
decay channel for ICARUS.}
\label{tab:icarus}
\begin{tabular}{@{}ccccccc}
\hline
 channel & \multicolumn{3}{c}{signal  for $\Delta m^{2}$  ($\rm{eV}^{2}$)}  &  $\epsilon$xBr & Background\\
       & 1.6x$10^{-3}$   & 2.5x$10^{-3}$ & 4.0x$10^{-3}$   &  &  \\
 \hline
$\tau\rightarrow$e           &  3.7 & 9.0 & 23 & 4.4\%  & 0.7 \\
$\tau\rightarrow$$\rho$ DIS  &  0.6 & 1.5 & 3.9 & 0.8\% & $<$0.1 \\
$\tau\rightarrow$$\rho$ QE   &  0.6 & 1.4 & 3.9 & 0.7\% & $<$0.1 \\
\hline
Total       &  4.9 & 11.9 & 30.5 & 5.9\% & 0.7 \\
\hline
\end{tabular}
\end{table*}

\section{Search for $\numunue$ appearance}
\begin{figure}[htb]
\vspace*{-0.2cm}
\begin{center}
    \includegraphics[width=7.5cm]{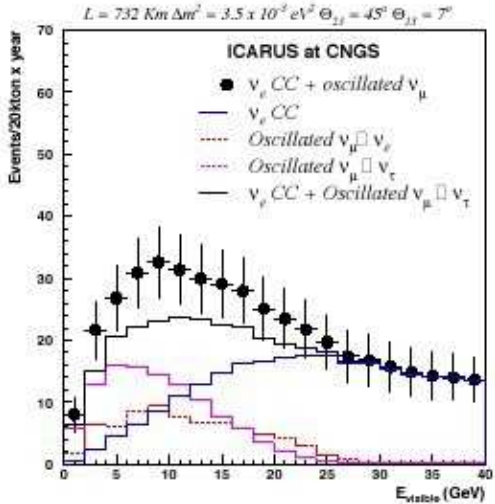}
\end{center}
\vspace{-1.5cm}
\caption{ The visible energy
distribution for the events contributing in the $\nu_{e}$ CC sample in ICARUS.}
\label{fig:evis}
\vspace{-0.7cm}
\end{figure}

In addition to the dominant $\numutau$ oscillation, it is possible that a 
sub-leading transition involving $\nu_{e}$ occurs as well. In the 3 flavour 
neutrino oscillation framework, assuming $\Delta m_{12}^{2}<<\Delta m_{23}^{2}
=\Delta m_{13}^{2}= \Delta m^{2}$, oscillation probabilities can be 
expressed like:
\begin{eqnarray*}
P(\numutau) = cos^{4}\theta_{13}sin^{2}2\theta_{23}sin^{2}(1.27\Delta m^{2}L/E)\\
P(\numunue) = sin^{2}\theta_{23}sin^{2}2\theta_{13}sin^{2}(1.27\Delta m^{2}L/E)
\end{eqnarray*}
 The sub-leading $\numunue$ oscillation at the
atmospheric scale is driven by the mixing angle $\theta_{13}$ which is constrained by
CHOOZ experiment to be small ($sin^{2}\theta_{13} <$0.14)~\cite{chooz}.
Having excellent electron identification capabilities,
both ICARUS and OPERA have estimated their sensitivity 
in searching for $\numunue$ appearance with the CNGS beam.
The analysis principle is based on a search for an excess of $\nu_{e}$ CC events at low
neutrino energies. The main background comes from the electron neutrino contamination 
present in the beam.   
The analysis takes into account the electron events coming from $\numutau$
events where $\tau\rightarrow e\nu_{\tau}\nu_{e}$ since both oscillations would occur
at the atmospheric $\Delta m^{2}$ 
scale. These events distort the kinematical distributions where
the low energy events contribute. This is illustrated in Fig.~\ref{fig:evis} showing
 the ICARUS visible energy
distribution for the events contributing in the $\nu_{e}$ CC sample
when $\theta_{13}=7^{o}$. The contribution
from both oscillations have different shapes than the rest of the events.

 The sensitivity to $\theta_{13}$ is obtained by doing a $\chi^{2}$ minimisation using
the visible energy, the missing transverse energy and the electron transverse momentum
distributions in which the oscillation parameters are allowed to vary. 
Table~\ref{tab:nue} summarises the expected number of selected  events for ICARUS assuming
5 years running (2.25x$10^{20}$ pot) from $\numunue$  and $\numutau$ oscillations in three family 
mixing and from background $\nu_{e}$ at 3 values of $\theta_{13}$.
\begin{table}[hbt]
\vspace*{-0.5cm}
\center
\caption{Expected number of signal and background  events in 5 years 
obtained in the search of $\numunue$ oscillation for ICARUS for 3 values
of $\theta_{13}$}
\label{tab:nue}
\begin{tabular}{@{}cccccc}
\hline
 $\theta_{13}$ & $sin^{2}2\theta_{13}$ & $\nu_{e}$CC  &  $\numutau$ & signal\\
  (deg) & & & $\tau\rightarrow$e& $\numunue$\\
 \hline

9 & 0.095 & 50 & 24 & 27 \\
7 & 0.058 & 50 & 24 & 16 \\
5 & 0.030 & 50 & 25 & 8.4 \\
\hline
\end{tabular}
\vspace*{-0.3cm}
\end{table}
The  limit obtained by ICARUS 
 at 90\% CL on $\theta_{13}$ is $5.8^{o}$ after 5 years.

The recent analysis performed by OPERA and described in details in Ref.~\cite{opnumunue}
 gives a limit of $7.1^{o}$ at 90\% CL. Both experimental results lead to significant
improvement over the actual CHOOZ limit and open an important window on the third
mixing angle.

\section{Conclusion}
The CNGS construction is progressing well. The project is on schedule and a startup is 
expected for june 2006. At the same time OPERA enters the construction phase and should
be ready to take data by 2006. ICARUS has successfully demonstrated its principle with
the full scale 300 ton  prototype technical run
 and the collaboration aims to build a 3000 ton detector by 2006. \\
The detector performances are such that an unambiguous $\numutau$ appearance
signal should be seen after only a few years of data taking.
Combining these observations, the two experiments expect
to see 20-25 $\tau$ events after 5 years with very little background at 
$\Delta m^{2}$ = 2.5x$10^{-3}$ $eV^{2}$.
 They can achieve a measurement of $\Delta m^{2}$ with
10\% accuracy. 
The very good electron identification and measurement of the two detectors give
the possibility to
explore the $\numunue$ appearance channel pushing down the $\theta_{13}$ limit below
$7^{o}$. This may correspond to  the best sensitivity that can be reached
before the JHF program~\cite{jhf} turns on.

\end{document}